\documentstyle[12pt]{article}
\newcommand{\nc}{\newcommand}
\nc{\rnc}{\renewcommand}
\rnc{\title}[1]{\Large\bf\mbox{}\\ \mbox{}\\#1\bigskip\medskip\\}
\rnc{\author}[1]{\large #1\\ \smallskip}
\nc{\address}[1]{{\narrower\normalsize\it #1\\}\medskip}
\nc{\e}[1]{{\em #1\/}}
\nc{\comment}[1]{}
\rnc{\thesection}{\arabic{section}\,.}
\rnc{\thesubsection}{\arabic{section}.\arabic{subsection}}
\rnc{\theequation}{\arabic{section}.\arabic{equation}}
\nc{\sect}[1]{\section{#1}\setcounter{equation}{0}}
\nc{\sub}[1]{\subsection{#1}}
\nc{\subsub}[1]{\subsubsection{#1}}
\nc{\beq}{\begin{equation}}
\nc{\beqa}{\begin{eqnarray}}
\nc{\eql}[1]{\label{Eqn#1}}
\nc{\bleq}[1]{\beq\eql{#1}}
\nc{\eeq}{\end{equation}}
\nc{\eeqa}{\end{eqnarray}}
\nc{\noeqno}{\nonumber\\}
\nc{\eqref}[1]{(\ref{Eqn#1})}
\nc{\sm}[1]{{\scriptstyle #1}}
\nc{\ssm}[1]{{\scriptscriptstyle #1}}
\nc{\pl}{\!+\!}
\nc{\mi}{\!-\!}
\nc{\W}[5]{W\!\left(\,\begin{array}{@{}cc|}#4&#3\\#1&#2\end{array}
\;#5\right)}
\nc{\WW}[5]{\overline{W}\!\!\left(\,\begin{array}{@{}cc|}#4&#3\\#1&#2\end{array}
\;#5\right)}
\rnc{\AA}{\bar{A}}
\nc{\Wf}[6]{W^{#1}\!\!\left(\,\begin{array}{@{}cc|}#5&#4\\#2&#3
\end{array}\;#6\right)}
\nc{\B}[4]{B\!\!\left(\left.\!#2\,
\begin{array}{c}#3\\#1\end{array}\!\right|#4\right)}
\nc{\BD}[3]{\bar{B}_{#1}(\,#2\mid#3\,)}
\nc{\BDD}[3]{\bar{B}'_{#1}(\,#2\mid#3\,)}
\rnc{\a}{\bar{a}}
\rnc{\b}{\bar{b}}
\rnc{\c}{\chi}
\nc{\ep}{\epsilon}
\nc{\cc}{\bar{\chi}}
\nc{\x}{\xi}
\nc{\xx}{\bar{\xi}}
\rnc{\l}{\lambda}
\nc{\m}{\mu}
\rnc{\t}{\theta}
\rnc{\vec}[1]{\mbox{\boldmath$#1$}}
\nc{\pos}[2]{\makebox(0,0)[#1]{$#2$}}
\nc{\spos}[2]{\makebox(0,0)[#1]{$\sm{#2}$}}
\nc{\text}[6]{\begin{picture}(#1,#2)
\put(#3,#4){\pos{#5}{\displaystyle#6}}\end{picture}}
\nc{\dl}[3]{\put(#1,#2){\makebox(#3,0){\dotfill}}}
\rnc{\d}[2]{\put(#1,#2){\spos{}{\bullet}}}
\nc{\dd}[3]{\multiput(#1,#2)(0,1){#3}{\spos{}{\bullet}}}
\begin{document}
\begin{center}
\title{A Construction of Solutions to Reflection Equations\\
for Interaction-Round-a-Face Models}
\author{Roger E. Behrend\footnote{E-mail: reb@maths.mu.oz.au}}
\address{Department of Mathematics, University of Melbourne\\Parkville,
Victoria 3052, Australia}
and\\
\medskip
\author{Paul A. Pearce\footnote{E-mail: pap@maths.mu.oz.au.
On leave from University of Melbourne.}}
\address{Physikalische Institut der Universit\"{a}t Bonn\\
Nu\ss allee 12, D-53115 Bonn, Germany}

\begin{abstract}
\noindent We present a procedure in which known solutions to
reflection equations for interaction-round-a-face lattice models are used
to construct  new solutions. The procedure is particularly well-suited
to models which have a known fusion hierarchy and which are based on graphs
containing a node of valency $1$.  Among such models are the
Andrews-Baxter-Forrester models, for which we construct reflection equation
solutions for fixed and free boundary conditions.
\end{abstract}
\end{center}
\sect{Introduction}
Boundary weights which satisfy reflection equations are important
in the study of solvable interaction-round-a-face (IRF) lattice models with
non-periodic boundary conditions~\cite{BehPeaObr95}--\cite{BatFriZho95}.
More specifically, such boundary weights lead to families of commuting
transfer matrices and hence integrability.
In~\cite{BehPeaObr95,AhnKoo95,BatFriZho95}, boundary weights
were obtained by directly
solving the IRF reflection equations, while in ~\cite{FanHouShi95}
they were obtained using intertwiners together with
known boundary weights for a related vertex model.

Here, we present a procedure in which known boundary weights for an
IRF model---together with auxiliary face weights, generally
obtained from a fusion hierarchy---are used to construct new
boundary weights for that model.
This procedure takes two forms, one which leads to weights for
fixed boundary conditions and the other which leads to weights for
free, or at least quasi-free, boundary conditions.  In each case,
the resulting boundary weights contain an arbitrary parameter.

Our procedure is particularly effective for models,
such as the Andrews-Baxter-Forrester (ABF) models~\cite{AndBaxFor84},
which are based on graphs containing a node of valency $1$, since there
then exist  trivial weights which can be used as the known, starting weights.
In this paper, we apply our procedure to the
ABF models and obtain weights for fixed boundary conditions which match
those of~\cite{BehPeaObr95}, as well as weights for free boundary conditions.

\sect{General Procedure}
We are considering an IRF model on a square lattice,
and we assume that there are restrictions on the spins allowed on any
adjacent lattice sites, as specified by an adjacency matrix
\[A_{ab}=\left\{\begin{array}{ll}0\;,&\mbox{spins $a$ and $b$ may not be
adjacent}\\
1\;,&\mbox{spins $a$ and $b$ may be adjacent}\end{array}\right.\]
For such models, we associate a Boltzmann weight with each set of spins
$a$, $b$, $c$, $d$ that are allowed to be adjacent around a face---\e{ie}
for which $A_{ab}\:A_{bc}\:A_{cd}\:A_{da}=1$. These weights are denoted
%
%
\setlength{\unitlength}{13mm}
\beq\raisebox{-0.8\unitlength}[0.8\unitlength][
0.8\unitlength]{\text{2}{1.6}{1}{0.8}{}{\W{a}{b}{c}{d}{u}}
\text{1}{1.6}{0.5}{0.8}{}{=}
\begin{picture}(1.6,1.6)
\multiput(0.3,0.8)(0.5,0.5){2}{\line(1,-1){0.5}}
\multiput(0.3,0.8)(0.5,-0.5){2}{\line(1,1){0.5}}
\put(0.24,0.8){\spos{r}{a}}\put(0.8,0.24){\spos{t}{b}}
\put(1.36,0.8){\spos{l}{c}}\put(0.8,1.36){\spos{b}{d}}
\put(0.8,0.8){\spos{}{u}}\end{picture}}\eql{FW}\eeq
where $u$ is the spectral parameter.

\sub{Fixed Boundary Conditions}
We now consider a boundary containing a fixed spin $\a$.
In this case we associate a boundary weight with each spin $a$ which
is allowed to be adjacent to $\a$
%
%
\setlength{\unitlength}{13mm}
\beq\raisebox{-0.5\unitlength}[0.5\unitlength][
0.5\unitlength]{\text{1.4}{1}{0.7}{0.5}{}{\BD{\a}{a}{u}}
\text{1}{1}{0.5}{0.5}{}{=}
\begin{picture}(1.4,1)
\put(0.3,0.5){\line(1,0){0.8}}
\put(0.24,0.5){\spos{r}{a}}\put(1.16,0.5){\spos{l}{\a}}
\put(0.7,0.58){\spos{b}{u}}
\multiput(1.1,0)(0,0.15){7}{\line(0,1){0.1}}
\end{picture}}\eql{DBW}\eeq
It can be shown~\cite{BehPeaObr95} that families of commuting transfer matrices
can be obtained if the boundary weights~\eqref{DBW}, together with the face
weights~\eqref{FW}, satisfy the fixed-boundary reflection equations
for $\a$. There is one such equation for each set of spins
$b$, $c$, $d$ satisfying $A_{\a b}\:A_{bc}\:A_{cd}\:A_{d\a}=1$,
%
%
\beqa\lefteqn{\rule[-3.5ex]{0ex}{3.5ex}
\sum_{f}\;\W{b}{\a}{f}{c}{u\!-\!v}\;
\W{c}{f}{\a}{d}{\m\!-\!u\!-\!v}\;\BD{\a}{f}{u}\;\BD{\a}{d}{v}=}
\qquad\noeqno
&&\sum_{f}\;\W{d}{c}{f}{\a}{u\!-\!v}\;\W{c}{b}{\a}{f}{\m\!-\!u\!-\!v}\;
\BD{\a}{f}{u}\;\BD{\a}{b}{v}\eql{DRE}\eeqa
\setlength{\unitlength}{18mm}\begin{center}
\begin{picture}(2.5,2)
\put(0.5,0){\begin{picture}(1.5,2)
\multiput(0,0.5)(0.5,-0.5){2}{\line(1,1){1}}
\multiput(0,0.5)(0.5,0.5){3}{\line(1,-1){0.5}}
\multiput(1,0.5)(0,1){2}{\line(1,0){0.5}}
\put(1.28,0.53){\spos{bl}{u}}\put(1.28,1.53){\spos{bl}{v}}
\put(0.5,0.5){\spos{}{u-v}}\put(1,1){\spos{}{\m-u-v}}
\multiput(1.5,0)(0,0.075){27}{\line(0,1){0.05}}\end{picture}}
\put(1,-0.02){\spos{t}{\a}}\multiput(2.04,0.5)(0,0.5){3}{\spos{l}{\a}}
\put(0.48,0.5){\spos{r}{b}}\put(0.98,1.02){\spos{br}{c}}
\put(1.48,1.52){\spos{br}{d}}\put(1.48,0.43){\spos{t}{f}}
\put(1.5,0.5){\spos{}{\bullet}}
\end{picture}
\text{0.4}{2}{0.2}{0.95}{}{=}
\begin{picture}(2.5,2)\put(0.5,0){\begin{picture}(1.5,2)
\multiput(1,0.5)(0.5,0.5){2}{\line(-1,1){1}}
\multiput(1,0.5)(-0.5,0.5){3}{\line(1,1){0.5}}
\multiput(1,0.5)(0,1){2}{\line(1,0){0.5}}
\put(1.28,0.53){\spos{bl}{v}}\put(1.28,1.53){\spos{bl}{u}}
\put(0.5,1.5){\spos{}{u-v}}\put(1,1){\spos{}{\m-u-v}}
\multiput(1.5,0)(0,0.075){27}{\line(0,1){0.05}}\end{picture}}
\put(1,2.02){\spos{b}{\a}}\multiput(2.04,0.5)(0,0.5){3}{\spos{l}{\a}}
\put(1.48,0.48){\spos{tr}{b}}\put(0.98,0.98){\spos{tr}{c}}
\put(0.48,1.5){\spos{r}{d}}\put(1.52,1.6){\spos{b}{f}}
\put(1.5,1.5){\spos{}{\bullet}}
\end{picture}\end{center}
Here, $\m$ is an arbitrary fixed parameter and the sums are over all
spins $f$ which are allowed to be adjacent to $\a$.
We note that if the face weights satisfy the symmetry
\beq\W{a}{b}{c}{d}{u}=\W{a}{d}{c}{b}{u}\eql{refsym}\eeq
then~\eqref{DRE} is automatically satisfied whenever $b=d$.
Furthermore, in the case in which there is only
one spin $a$ allowed to be adjacent to $\a$---\e{ie} $\a$ has a
valency of $1$---we must have $b=d=f=a$ in~\eqref{DRE}
implying that the equation is always satisfied and that
the single boundary weight $\BD{\a}{a}{u}$ may be assigned to any
function of $u$.

\sub{Free Boundary Conditions}
For the case of free, or at least quasi-free, boundary conditions,
we associate a boundary weight with each
set of spins $a$, $b$, $c$ satisfying $A_{ab}\:A_{bc}=1$,
%
%
\setlength{\unitlength}{13mm}
\beq\raisebox{-0.8\unitlength}[0.8\unitlength][
0.8\unitlength]{\text{1.6}{1.6}{0.8}{0.8}{}{\B{a}{b}{c}{u}}
\text{1}{1.6}{0.5}{0.8}{}{=}
\begin{picture}(1.1,1.6)
\multiput(0.8,0.3)(0,0.15){7}{\line(0,1){0.1}}
\put(0.8,0.3){\line(-1,1){0.5}}\put(0.8,1.3){\line(-1,-1){0.5}}
\put(0.84,0.26){\spos{tl}{a}}\put(0.24,0.8){\spos{r}{b}}
\put(0.84,1.34){\spos{bl}{c}}
\put(0.72,0.8){\spos{r}{u}}\end{picture}}
\eql{BW}\eeq
In this case, families of commuting transfer matrices can be obtained if
the boundary weights~\eqref{BW}, together with the face weights~\eqref{FW},
satisfy the free-boundary reflection equations.
There is one such equation for each set of spins $a$, $b$, $c$, $d$,
$e$ satisfying $A_{ab}\:A_{bc}\:A_{cd}\:A_{de}=1$,
%
%
\beqa\lefteqn{\rule[-3.5ex]{0ex}{3.5ex}\sum_{f
g}\;\W{b}{a}{f}{c}{u\!-\!v}\;
\W{c}{f}{g}{d}{\m\!-\!u\!-\!v}\;\B{a}{f}{g}{u}\;\B{g}{d}{e}{v}=}
\qquad\noeqno
&&\sum_{fg}\;\W{d}{c}{f}{e}{u\!-\!v}\;\W{c}{b}{g}{f}{\m\!-\!u\!-\!v}\;
\B{g}{f}{e}{u}\;\B{a}{b}{g}{v}\eql{RE}\eeqa
\setlength{\unitlength}{18mm}\begin{center}
\begin{picture}(2.5,2)
\put(0.5,0){\begin{picture}(1.5,2)
\multiput(1.5,0)(0,0.075){27}{\line(0,1){0.05}}
\put(1.5,1){\line(-1,1){0.5}}
\put(1.5,1){\line(-1,-1){1}}\put(1.5,0){\line(-1,1){1}}
\put(1.5,2){\line(-1,-1){1.5}}\put(0,0.5){\line(1,-1){0.5}}
\put(1.4,0.5){\spos{r}{u\,}}\put(1.4,1.5){\spos{r}{v\,}}
\put(0.5,0.5){\spos{}{u-v}}
\put(1,1){\spos{}{\m-u-v}}\end{picture}}
\put(2.03,2){\spos{l}{e}}\put(1.48,1.52){\spos{br}{d}}
\put(0.98,1.02){\spos{br}{c}}\put(0.48,0.5){\spos{r}{b}}
\put(1,-0.02){\spos{t}{a}}\put(2.03,0){\spos{l}{a}}
\put(1.48,0.43){\spos{t}{f}}\put(2.06,1){\spos{l}{g}}
\put(1.5,0.5){\spos{}{\bullet}}\put(2,1){\spos{}{\bullet}}
\end{picture}
\text{0.4}{2}{0.2}{0.95}{}{=}
\begin{picture}(2.5,2)\put(0.5,0){\begin{picture}(1.5,2)
\multiput(1.5,0)(0,0.075){27}{\line(0,1){0.05}}
\put(1.5,1){\line(-1,1){1}}\put(1.5,1){\line(-1,-1){0.5}}
\put(1.5,0){\line(-1,1){1.5}}\put(1.5,2){\line(-1,-1){1}}
\put(0,1.5){\line(1,1){0.5}}
\put(1.4,0.5){\spos{r}{v\,}}\put(1.4,1.5){\spos{r}{u\,}}
\put(1,1){\spos{}{\m-u-v}}\put(0.5,1.5){\spos{}{u-v}}\end{picture}}
\put(2.03,2){\spos{l}{e}}\put(1,2.02){\spos{b}{e}}
\put(0.48,1.5){\spos{r}{d}}
\put(0.98,0.98){\spos{tr}{c}}\put(1.48,0.48){\spos{tr}{b}}
\put(2.03,0){\spos{l}{a}}\put(1.52,1.6){\spos{b}{f}}
\put(2.06,1){\spos{l}{g}}
\put(1.5,1.5){\spos{}{\bullet}}\put(2,1){\spos{}{\bullet}}
\end{picture}\end{center}
\rule{0ex}{3ex}Here, the sum on the left side is over all spins $f$, $g$
satisfying $A_{af}\:A_{cf}\:A_{fg}\:A_{gd}=1$ and that on the right
side is over all spins $f$, $g$ satisfying
$A_{ef}\:A_{cf}\:A_{fg}\:A_{gb}=1$.

\sub{Construction of New Boundary Weights}
Our construction of new boundary weights requires that there
exist an auxiliary adjacency matrix $\AA$ and,
for each set of spins $a$, $b$, $c$, $d$ satisfying
$\AA_{ab}\:A_{bc}\:\AA_{cd}\:A_{da}=1$, an
auxiliary face weight
%
%
\setlength{\unitlength}{8mm}
\beq\raisebox{-1\unitlength}[1\unitlength][
1\unitlength]{\text{3.2}{2}{1.6}{1}{}{\WW{a}{b}{c}{d}{u}}
\text{1.6}{2}{0.8}{1}{}{=}
\begin{picture}(3,2)
\multiput(0.5,0.5)(2,0){2}{\line(0,1){1}}
\multiput(0.5,0.5)(0,1){2}{\line(1,0){2}}
\put(0.45,0.45){\spos{tr}{a}}\put(2.55,0.45){\spos{tl}{b}}
\put(2.55,1.55){\spos{bl}{c}}\put(0.45,1.55){\spos{br}{d}}
\put(1.5,1){\spos{}{u}}\end{picture}}\eql{AFW}\eeq
These weights, together with the fundamental face weights~\eqref{FW},
are assumed to satisfy the generalised Yang-Baxter equations.  There is
one such equation for each set of spins $a$, $b$, $c$, $d$, $e$, $f$
satisfying $A_{ab}\:\AA_{bc}\:A_{cd}\:A_{de}\:\AA_{ef}\:A_{fa}=1$,
%
%
\beqa\lefteqn{\rule[-3.5ex]{0ex}{3.5ex}\sum_g\;
\W{a}{b}{g}{f}{u\mi v}\;\WW{b}{c}{d}{g}{u}\;
\WW{g}{d}{e}{f}{v}=}\qquad\qquad\qquad\qquad\noeqno
&&\sum_g\;\WW{b}{c}{g}{a}{v}\;\WW{a}{g}{e}{f}{u}\;
\W{g}{c}{d}{e}{u\mi v}\eql{GYBE}\eeqa
\setlength{\unitlength}{8mm}\begin{center}
\begin{picture}(5,3)
\multiput(0.5,1.5)(1,-1){2}{\line(1,1){1}}
\multiput(0.5,1.5)(1,1){2}{\line(1,-1){1}}
\multiput(2.5,0.5)(0,1){3}{\line(1,0){2}}
\multiput(2.5,0.5)(2,0){2}{\line(0,1){2}}
\put(1.5,1.5){\spos{}{u-v}}\put(3.5,1){\spos{}{u}}
\put(3.5,2){\spos{}{v}}
\put(0.4,1.5){\spos{r}{a}}\put(1.5,0.4){\spos{t}{b}}
\put(2.5,0.4){\spos{t}{b}}\put(4.55,0.45){\spos{tl}{c}}
\put(4.6,1.5){\spos{l}{d}}\put(4.55,2.55){\spos{bl}{e}}
\put(1.5,2.6){\spos{b}{f}}\put(2.5,2.6){\spos{b}{f}}
\put(2.58,1.42){\spos{tl}{g}}
\put(2.5,1.5){\spos{}{\bullet}}\end{picture}
\text{1}{3}{0.5}{1.5}{}{=}
\begin{picture}(5,3)
\multiput(2.5,1.5)(1,-1){2}{\line(1,1){1}}
\multiput(2.5,1.5)(1,1){2}{\line(1,-1){1}}
\multiput(0.5,0.5)(0,1){3}{\line(1,0){2}}
\multiput(0.5,0.5)(2,0){2}{\line(0,1){2}}
\put(3.5,1.5){\spos{}{u-v}}\put(1.5,1){\spos{}{v}}
\put(1.5,2){\spos{}{u}}
\put(0.4,1.5){\spos{r}{a}}\put(0.45,0.45){\spos{tr}{b}}
\put(2.5,0.4){\spos{t}{c}}\put(3.5,0.4){\spos{t}{c}}
\put(4.6,1.5){\spos{l}{d}}\put(2.5,2.6){\spos{b}{e}}
\put(3.5,2.6){\spos{b}{e}}\put(0.45,2.55){\spos{br}{f}}
\put(2.42,1.42){\spos{tr}{g}}
\put(2.5,1.5){\spos{}{\bullet}}\end{picture}\end{center}
Here, the sum on the left side is over all spins $g$
satisfying $A_{bg}\:A_{fg}\:\AA_{gd}=1$ and that on the right
side is over all spins $g$ satisfying $\AA_{ag}\:A_{gc}\:A_{ge}=1$.

In practice, the auxiliary face weights can generally be constructed
using fusion with an appropriate row of fundamental face weights.

Our construction of new boundary weights takes two forms.
In the first form, we obtain new weights for a
boundary with fixed spin $\a$ using known weights for a
boundary with fixed spin $\b$, where we assume that, with respect
to $\AA$, $\a$ is the only spin allowed to be adjacent to $\b$.
The new weights depend on an arbitrary parameter $\cc$ and,
for each spin $a$ allowed to be adjacent to $\a$, are defined as
%
%
\setlength{\unitlength}{8mm}
\beqa\BDD{\a}{a}{u}&=&\sum_{b}\;\WW{\a}{\b}{b}{a}{u\pl\cc}\;
\WW{a}{b}{\b}{\a}{\m\mi u\pl\cc}\;\BD{\b}{b}{u}\eql{NDBW}\\
&\text{1}{3.2}{0.5}{1.5}{}{=}&\begin{picture}(4,3)
\multiput(0.5,0.5)(0,2){2}{\line(1,0){2}}
\multiput(0.5,0.5)(2,0){2}{\line(0,1){2}}
\put(0.5,1.5){\line(1,0){3}}
\multiput(3.5,0.5)(0,0.3){7}{\line(0,1){0.2}}
\put(3,1.56){\spos{b}{u}}
\put(1.5,1){\spos{}{u+\cc}}\put(1.5,2){\spos{}{\m-u+\cc}}
\put(0.4,1.5){\spos{r}{a}}\put(3.6,1.5){\spos{l}{\b}}
\put(2.5,0.4){\spos{t}{\b}}\put(2.5,2.6){\spos{b}{\b}}
\put(0.45,0.45){\spos{tr}{\a}}\put(0.45,2.55){\spos{br}{\a}}
\put(2.42,1.42){\spos{tr}{b}}\put(2.5,1.5){\spos{}{\bullet}}
\end{picture}\nonumber\eeqa
where the sum is over all spins $b$ satisfying $\AA_{ab}\:A_{b\b}=1$.
We note that we suppress the dependence of these weights on the
the spin $\b$ and the parameter $\cc$.
It is straightforward to show that
the new weights~\eqref{NDBW} satisfy the fixed boundary reflection
equations for $\a$, using
the assumptions that the known weights satisfy the fixed
boundary reflection equations for $\b$,
that the auxiliary face weights satisfy~\eqref{GYBE},
and that $\b$ has valency $1$ with respect to $\AA$.

In the second form of our construction of new boundary weights,
we obtain certain weights for free boundary conditions using known weights
for a boundary with fixed spin $\a$.
The new weights depend on an arbitrary parameter $\c$ and,
for each set of spins $a$, $b$, $c$ satisfying
$\AA_{\a a}\:A_{ab}\:A_{bc}\:\AA_{c\a}=1$, are defined as
%
%
\setlength{\unitlength}{8mm}
\beqa\B{a}{b}{c}{u}&=&\sum_d\;\WW{a}{\a}{d}{b}{u\pl\c}\;
\WW{b}{d}{\a}{c}{\m\mi u\pl\c}\;\BD{\a}{d}{u}\eql{NBW}\\
&\text{1}{3.2}{0.5}{1.5}{}{=}&\begin{picture}(4,3)
\multiput(0.5,0.5)(0,2){2}{\line(1,0){2}}
\multiput(0.5,0.5)(2,0){2}{\line(0,1){2}}
\put(0.5,1.5){\line(1,0){3}}
\multiput(3.5,0.5)(0,0.3){7}{\line(0,1){0.2}}
\put(3,1.56){\spos{b}{u}}
\put(1.5,1){\spos{}{u+\chi}}\put(1.5,2){\spos{}{\m-u+\c}}
\put(0.4,1.5){\spos{r}{b}}\put(3.6,1.5){\spos{l}{\a}}
\put(2.5,0.4){\spos{t}{\a}}\put(2.5,2.6){\spos{b}{\a}}
\put(0.45,0.45){\spos{tr}{a}}\put(0.45,2.55){\spos{br}{c}}
\put(2.42,1.42){\spos{tr}{d}}\put(2.5,1.5){\spos{}{\bullet}}
\end{picture}\nonumber\eeqa
where the sum is over all spins $d$ satisfying $\AA_{bd}\:A_{d\a}=1$.
Again we suppress the dependence of these weights on
the spin $\a$ and the parameter $\c$.
The new weights~\eqref{NBW} satisfy the free boundary reflection
equations for each set of spins $a$, $b$, $c$, $d$,
$e$ in~\eqref{RE} which satisfy
$\AA_{\a a}\:A_{ab}\:A_{bc}\:A_{cd}\:A_{de}\:\AA_{e\a}=1$ .
This follows straightforwardly
from the assumptions that the known weights satisfy the fixed
boundary reflection equations for $\a$,
and that the auxiliary face weights satisfy~\eqref{GYBE}.

\sect{ABF Models}
We now consider the Andrews-Baxter-Forrester (ABF)
models~\cite{AndBaxFor84}.
There is one such model for each integer $L\geq3$, with the spins $a$
in this model taking the values
\beq a\in\{1,2,\ldots,L\}\eeq
The adjacency matrix is defined by the condition that $A_{ab}=1$
if and only if
\beq |a-b|=1\eql{ABFAC}\eeq
There is a fixed crossing parameter
\beq\l=\frac{\pi}{L+1}\eql{lambda}\eeq
and the face weights are given by
\beqa\rule[-3.5ex]{0ex}{3.5ex}
\W{a}{a\!\mp\!1}{a}{a\!\pm\!1}{u}&=&
\frac{\t(\l\!-\!u)}{\t(\l)}\noeqno
\rule[-3.5ex]{0ex}{3.5ex}\W{a\!\mp\!1}{a}{a\!\pm\!1}{a}{u}&=&
\sqrt{\frac{\t((a\!-\!1)\l)\,\t((a\!+\!1)\l)}{\t(a\l)^2}}\;
\frac{\t(u)}{\t(\l)}\eql{ABFFW}\\
\W{a\!\pm\!1}{a}{a\!\pm\!1}{a}{u}&=&\frac{\t(a\l\!\pm\!u)}{\t(a\l)}
\nonumber\eeqa
where $\t$ is the standard elliptic theta-$1$ function of fixed nome.

For these models, an auxiliary adjacency matrix
and auxiliary face weights which satisfy~\eqref{GYBE}
are provided by the level $n$ fused adjacency matrix
and the $n$ by $1$ fused face
weights~\cite{DatJimMiwOka86,DatJimMiwOka87a,DatJimKunMiwOka88}
\beq\AA=A^n,\qquad\overline{W}=W^{n,1}\eql{AW}\eeq
where
\beq n\in\{0,1,\ldots,L\mi1\}\eeq
The level~$n$ fused adjacency matrix is defined by the condition that
$A^n_{ab}=1$ if and only if
\beq a-b\:\in\:\{-n,\:-n\!+\!2,\:\ldots\:,\:n\!-\!2,\:n\}\eql{AC1}\eeq
and
\beq a+b\:\in\:\{n\!+\!2,\:n\!+\!4,\:\ldots\:,\:2L\!-\!n\!-\!2,\:
2L\!-\!n\}\eql{AC2}\eeq
We note that $A^1=A$.
The $n$ by $1$ fused face weights are defined in terms
of rows of $n$ fundamental face weights~\eqref{ABFFW} and, after
appropriate normalisation and symmetrisation, are given by
\beqa\lefteqn{\Wf{n,1}{a}{b}{c}{d}{u}\;=}\qquad\qquad\qquad\eql{ABFFFW}\\
&&\left\{\begin{array}{ll}\rule[-5ex]{0ex}{5ex}
\displaystyle\ep_b\,\ep_d\;\sqrt{\frac{\t(\frac{a+b\mp n}{2}\l)\;
\t(\frac{c+d\pm n}{2}\l)}{\t(b\l)\;\t(d\l)}}\;\,
\frac{\t(u\mi\frac{n\pm(a-b)}{2}\l)}{\t(\l)}&;\;
c=b\!\pm\!1,\;d=a\!\pm\!1\\
\displaystyle\ep_b\,\ep_d\;\sqrt{\frac{\t(\frac{n\mp(a-b)}{2}\l)\;
\t(\frac{n\pm(d-c)}{2}\l)}{\t(b\l)\;\t(d\l)}}\;\,
\frac{\t(\frac{a+b\pm n}{2}\l\!\mp\!u)}{\t(\l)}&;\;
c=b\!\mp\!1,\;d=a\!\pm\!1\end{array}\right.\nonumber\eeqa
where $\ep_a$ are factors whose required properties are
\beq(\ep_a)^2=1\;,\qquad\qquad\ep_a\;\ep_{a\!+\!2}=-1\eeq
We note that the fused weights~\eqref{ABFFFW}
reduce to~\eqref{ABFFW} for $n=1$.

\sub{Weights for Fixed Boundary Conditions}
Since, for the ABF models, the spin $1$ has valency $1$
with respect to $A$, and the face weights satisfy the
symmetry~\eqref{refsym}, the boundary weight $\BD{1}{2}{u}$
can be set to an arbitrary function of $u$.
Furthermore, it follows from~\eqref{AC1} and~\eqref{AC2} that
the spin $1$ has valency $1$ with respect to any $A^{\a\mi1}$,
the only allowed neighbour being the spin $\a$.  It is therefore
possible to construct new weights for a boundary
with fixed spin $\a$ using an arbitrary weight for a boundary with
fixed spin $1$.  Accordingly, we apply~\eqref{NDBW} with
$\AA=A^{\a\mi1}$, $\overline{W}=W^{\a\mi1,1}$, $\b=1$,
$\m=\l$, $\cc=-\!\l\mi\xx$ and
$\BD{1}{2}{u}=\ep_1\,\ep_2\,\ep_{\a}\,\ep_{\a\mi1}\,
\sqrt{\t(2\l)/\t(\l)}\,g(u)$, which gives
\beq\BDD{\a}{\a\!\pm\!1}{u}=g(u)\;
\sqrt{\frac{\t((\a\!\pm\!1)\l)}{\t(\a\l)}}\;\;\frac{\t(u\!\pm\!\xx)
\,\t(u\!\mp\!\a\l\!\mp\!\xx)}{\t(\l)^2}\eql{ABFDBW}\eeq
where $\xx$ is an arbitrary constant and $g$ is an arbitrary function.
It can be seen that these weights exactly match those obtained
in~\cite{BehPeaObr95} by directly solving the reflection equations.

\sub{Weights for Free Boundary Conditions}
We now consider the construction of ABF weights for free boundary
conditions using~\eqref{NBW}\rule[-2ex]{0ex}{2ex} together with~\eqref{AW}
and~\eqref{ABFDBW}.
We shall associate with any ABF weight $\B{a}{b}{c}{u}$
either
odd\rule{0ex}{3ex} or even parity, according to the parity of $b$.
Due to~\eqref{ABFAC}, each free boundary reflection equation~\eqref{RE}
contains boundary weights all with the same parity.
Similarly, due to \eqref{AC1} and~\eqref{AC2},~\eqref{NBW}
generates boundary weights all with the same parity.
The requirement in~\eqref{NBW} that we have
$A^n_{a\a}\:A^n_{c\a}=1$ also implies that,
in general, there might not be a \rule[-2ex]{0ex}{2ex}weight
$\B{a}{b}{c}{u}$ generated
for each $b$ of the appropriate parity.
However, by examining~\eqref{AC1} and~\eqref{AC2}\rule{0ex}{3ex},
we find that the unique values
\beq n=\left\{\begin{array}{ll}
\rule[-2ex]{0ex}{2ex}\frac{L-1}{2}&\mbox{, $L$ odd}\\
\frac{L}{2}&\mbox{, $L$ even}
\end{array}\right.\qquad\qquad\qquad
\a=\left\{\begin{array}{ll}
n&\mbox{, odd weights}\\
n+1&\mbox{, even weights}
\end{array}\right.\eql{an}\eeq
do generate a full set of boundary weights of a given parity.

We now apply~\eqref{NBW} with $\m=\l$, $\c=\x\pl\frac{n-1}{2}\l$,
$g(u)\mapsto\ep_{\a}\,\ep_{\a\mi1}\,g(u)$
and $\xx\mapsto\xx\mi\frac{\a}{2}\l$, which gives
\beqa\rule[-5ex]{0ex}{5ex}\lefteqn{\B{a\!\mp\!1}{a}{a\!\pm\!1}{u}\;=\;
g(u)\;\sqrt{\frac{\t(a\l)}{\t((a\!\pm\!1)\l)}}\quad\times}\eql{ABFNDBW1}\\
&&\sqrt{\frac{\t(\frac{a+\a-n-1}{2}\l)\;
\t(\frac{a+\a+n+1}{2}\l)\;
\t(\frac{n+1-a+\a}{2}\l)\;
\t(\frac{n+1+a-\a}{2}\l)}{\t(a\l)^2\:\t(\l)^2}}\;\;
\frac{\t(\frac{a}{2}\l\mi\xx\!\mp\!\x)\;
\t(\frac{a}{2}\l\pl \xx\!\mp\!\x)}{\t(\l)^2}\;\;
\frac{\t(2u)}{\t(\l)}\nonumber\eeqa
\beqa\rule[-5ex]{0ex}{5ex}\lefteqn{\B{a\!\pm\!1}{a}{a\!\pm\!1}{u}\;=\;
g(u)\;\sqrt{\frac{\t(a\l)}{\t((a\!\pm\!1)\l)}}\quad\times}\eql{ABFNDBW2}\\
\lefteqn{\rule[-5ex]{0ex}{5ex}
\quad\left(\frac{\t(\frac{n+1-a+\a}{2}\l)\;
\t(\frac{n+1+a-\a}{2}\l)\;
\t(u\!\pm\!\frac{a+\a}{2}\l\mi\x)\;
\t(u\!\pm\!\frac{a+\a}{2}\l\pl\x)\;
\t(u\!\mp\!\frac{\a}{2}\l\mi\xx)\;
\t(u\!\mp\!\frac{\a}{2}\l\pl\xx)}{\t(a\l)\:\t(\a\l)\:\t(\l)^4}\;\;
+\right.}\noeqno
&&\qquad\left.\frac{\t(\frac{a+\a-n-1}{2}\l)\;
\t(\frac{a+\a+n+1}{2}\l)\;
\t(u\!\pm\!\frac{a-\a}{2}\l\mi\x)\;
\t(u\!\pm\!\frac{a-\a}{2}\l\pl\x)
\t(u\!\pm\!\frac{\a}{2}\l\mi\xx)\;
\t(u\!\pm\!\frac{\a}{2}\l\pl\xx)}{\t(a\l)\:\t(\a\l)\:\t(\l)^4}
\right)\nonumber\eeqa
Here, the two terms which led to~\eqref{ABFNDBW1} were combined using
a standard elliptic identity, and a common factor
$\ep_{a}\,\ep_{a\mi1}$ in~\eqref{ABFNDBW1} and~\eqref{ABFNDBW2}
was eliminated since, for a given $\a$, the allowed values of
$a$ must all have the same parity implying that this factor always
produces the same sign.

\sect{Discussion}
We have presented a general procedure for obtaining
boundary weights for IRF models and have applied this to the ABF models.
Our method should be useful for determining classes of IRF models
for which solutions of the reflection equations exist
and contain arbitrary parameters.  In particular,
our method implies the existence of such solutions for the
standard $A$-$D$-$E$ models, since these are all based on graphs
containing a node of valency $1$, and have known fusion hierarchies.

In future work, we plan to
construct the weights for the $D$ and $E$ series within the standard
$A$-$D$-$E$ models, and
to study further the weights obtained here for the ABF models,
which form the $A$ series. In particular, we intend to investigate the
relationship between the ABF weights for free boundary conditions found
using our method, and those obtained using intertwiners
or by directly solving the reflection equations.  We also
hope to be able to show that these weights
can be used to obtain genuine free boundary conditions at the
isotropic point, and that the associated transfer matrices
satisfy functional equations with the same form as in the case
of fixed and periodic boundary conditions.

\section*{Acknowledgements} PAP thanks Jean-Marie Maillard
and Jean-Bernard Zuber for their kind hospitality at Paris and
Saclay, where some of this work was done.  PAP also thanks Vladimir
Rittenberg for his kind hospitality at Bonn.
This work was supported by the Australian Research Council.

\end{document}